\documentclass[useAMS,usenatbib,letterpaper]{mn2e}
\usepackage[totalwidth=515pt, totalheight=680pt]{geometry}

\usepackage{graphicx}
\usepackage[fleqn]{amsmath}
\usepackage{amssymb}
\usepackage{color}

\newcommand{\cs}{c_\mathrm{s}}

\newcommand{\betac}{\beta_\mathrm{c}}

\title[Numerical convergence in self-gravitating disc simulations]{Numerical convergence in self-gravitating disc simulations: initial conditions and edge effects}

\author[S.-J. Paardekooper et al.]{Sijme-Jan Paardekooper$^1$\thanks{E-mail: 
S.Paardekooper@damtp.cam.ac.uk}
, Cl\'ement Baruteau$^1$ and Farzana Meru$^2$
\\  
$^1$DAMTP, University of Cambridge, Wilberforce Road, Cambridge CB3 0WA,
United Kingdom
\\$^2$Institut f\"ur Astronomie \& Astrophysik, Universit\"at T\"ubingen, Auf der Morgenstelle 10, T\"ubingen D-72076, Germany
}

\begin{document}

\date{Draft version \today}

\pagerange{\pageref{firstpage}--\pageref{lastpage}} \pubyear{2011}

\maketitle

\label{firstpage}

\begin{abstract}
We study the numerical convergence of hydrodynamical simulations of self-gravitating accretion discs, in which a simple cooling law is balanced by shock heating. It is well-known that there exists a critical cooling time scale for which shock heating can no longer compensate for the energy losses, at which point the disc fragments. The numerical convergence of previous results of this critical cooling time scale was questioned recently using Smoothed Particle Hydrodynamics (SPH). We employ a two-dimensional grid-based code to study this problem, and find that for smooth initial conditions, fragmentation is possible for slower cooling as the resolution is increased, in agreement with recent SPH results. We show that this non-convergence is at least partly due to the creation of a special location in the disc, the boundary between the turbulent and the laminar region, when cooling towards a gravito-turbulent state. Converged results appear to be obtained in setups where no such sharp edges appear, and we then find a critical cooling time scale of $\sim 4\Omega^{-1}$, where $\Omega$ is the local angular velocity.
\end{abstract}
 


\section{Introduction}
In the absence of any heating mechanism, a self-gravitating accretion disc will cool down until it becomes gravitationally unstable, which happens when the stability parameter $Q$ \citep{toomre64}
\begin{equation}
Q=\frac{\cs \kappa}{\pi G \Sigma}\approx 1
\end{equation}
Here, $\cs$ is the sound speed, $\kappa$ denotes the epicyclic frequency, and $\Sigma$ is the surface density of the disc. When $Q$ approaches unity and gravitational instabilities kick in, spiral shocks are generated that heat up the disc \citep{gold65,durisen07}.  An equilibrium can then be set up in which shock heating exactly balances energy losses through cooling \citep{pac78}. This is called the gravito-turbulent state of the disc. If the cooling time scale is parametrised as
\begin{equation}
t_\mathrm{cool}=\beta \Omega^{-1},
\label{eqtcool}
\end{equation}
with $\beta$ constant and $\Omega$ the local angular velocity of the disc, a local balance of heating and cooling dictates that the spiral shocks generate an effective $\alpha$ parameter \citep{pringle81,gammie01}
\begin{equation}
\alpha = \frac{4}{9} \frac{1}{\gamma(\gamma-1)\beta},
\label{eqabc}
\end{equation}
where $\gamma$ is the adiabatic exponent. A lot of work has been put into determining whether a local equilibrium really is established and whether an $\alpha$-description is valid at all for gravitoturbulent discs \citep{lodato04,lodato05,forgan11}, which is not necessarily the case \citep{balbus99}.

For strong enough cooling, the disc is unable to provide enough shock heating to balance the cooling, and the disc fragments \citep{gammie01}, possibly forming gas giant planets. The critical value of $\beta$ was found to be $\betac\approx 3$ in \cite{gammie01} for 2D simulations with $\gamma=2$. \cite{rice05} interpreted the fragmentation criterion as a maximum stress the disc can sustain. They found $\alpha_\mathrm{max}\approx 0.06$, which leads to a value of $\betac$ that depends on $\gamma$ through equation (\ref{eqabc}). It was found by \cite{clarke07} that the value of $\alpha_\mathrm{max}$ can increase by a factor of $\approx 2$ when $\beta$ is decreased slowly. 

\begin{figure*} 
\centering
\resizebox{0.9\hsize}{!}{\includegraphics[]{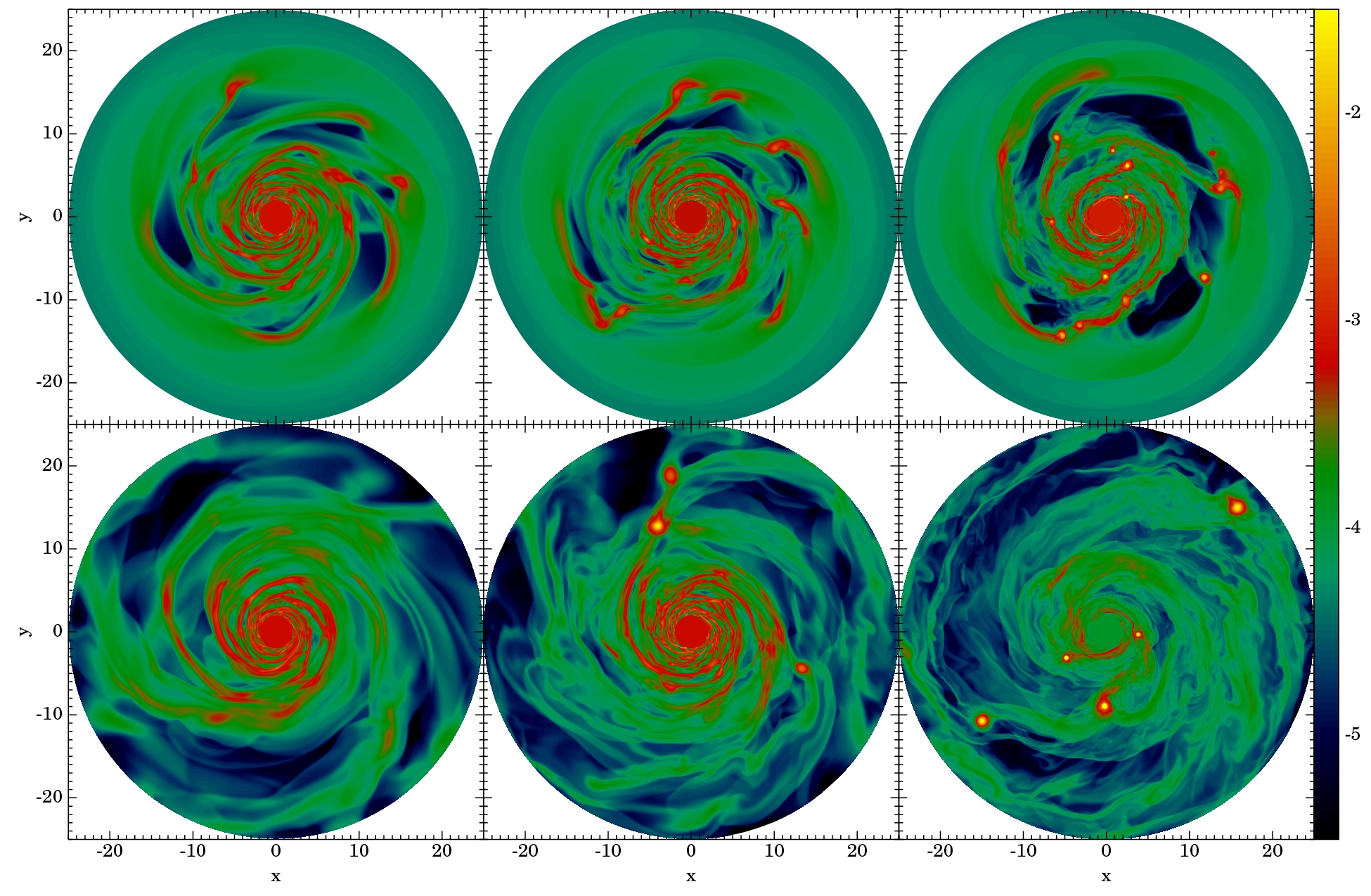}} 
\caption{Surface density after 150 (top row) and 300 (bottom row) orbits at $R=1$ for $\beta=7.5$. From left to right, the resolution is $\Delta R/R=0.008$, $0.004$ and $0.002$. At $R=10$, the scale height $H$ is resolved by approximately 10, 20 and 40 cells (from left to right).} 
\label{figdens} 
\end{figure*} 

The determination of the critical value of $\beta$ is important in models of giant planet formation that rely on disc instability \citep[e.g.][]{boss97,boley10}. Since according to the results above, the cooling time scale should be of the order of the orbital time scale, planet formation through disc instability is more likely to occur in the outer parts of real protoplanetary discs \citep{rafikov05,boley09}. 

More recently, the numerical convergence of the results on $\betac$ was questioned by \cite{meru11, meru11b}. There, it was shown through SPH simulations that the critical value of $\beta$ had not converged with respect to resolution. At higher and higher resolution, discs with higher and higher $\beta$ would fragment, with no sign of convergence. It was suggested by \cite{lodato11} that this behaviour might be due to the SPH artificial viscosity or the artificial smoothing of density enhancements. 

In this Letter, we present results on the convergence of the critical value of $\beta$ using the grid-based hydrodynamics code {\sc fargo} \citep[Fast Advection in Rotating Gaseous Objects,][see also section \ref{secNum}]{masset00a,masset00b}. First of all, with a smooth initial setup as in many previous simulations of gravitationally unstable discs, we find similar non-convergence with {\sc fargo}, indicating that the problem is not specific to SPH. We then show that numerical convergence can be reached by avoiding a sharp boundary between the gravito-turbulent inner part of the disc and the still laminar outer part, which arises when using smooth initial conditions during the initial stage of cooling towards $Q\approx 1$. If smooth initial conditions are not used, we find a value of $\betac$ close to that found by \cite{gammie01}.

\section{Numerical method}
\label{secNum}
We use the grid-based hydrodynamics code {\sc fargo} \citep{masset00a,masset00b}, for which a self-gravity solver was presented in \cite{baruteau08}. We have implemented the simple cooling law
\begin{equation}
\frac{d\epsilon}{dt}=-\frac{\epsilon}{t_\mathrm{cool}},
\end{equation}
where $\epsilon$ is the internal energy and $t_\mathrm{cool}$ is given by equation (\ref{eqtcool}). A standard prescription for artificial viscosity \citep{stone92} is used to handle shocks. Following \cite{lodato04}, we choose our disc to lie between $R=0.25$ and $R=25$. The self-gravity module of {\sc fargo} requires a logarithmic grid in the radial direction, and we choose our grid so as to give square cells everywhere ($\Delta R/R \approx \Delta \phi$). Our lowest resolution has 512 cells in the radial direction and 768 cells in the azimuthal direction, and we increase the resolution by a factor of 2 and 4. We use outflow boundary conditions at the inner and outer grid edges. 

Following \cite{lodato04}, we choose our disc to have surface mass density $\Sigma \propto R^{-1}$ and temperature $T \propto R^{-1/2}$. The angular velocity is Keplerian, corrected for the mass and pressure of the disc. A small level ($\sim 0.1\%$) of white noise is put on top to break the axisymmetry and allow spiral waves to form. The surface density and aspect ratio $H/R$ at $R=1$ are chosen so as to give a total disc mass of $0.25$ $M_*$, and so that the minimum value of $Q\approx 2$ at the outer edge of the disc. We use $\gamma=5/3$ throughout. 

We measure the total shear stress in the simulations by calculating
\begin{equation}
T_{R\varphi}=\left< \int  \frac{g_R g_\varphi}{4\pi G}dz\right> +\left<\Sigma \delta v_R \delta v_\varphi \right>,
\end{equation}
where $g_R$ and $g_\varphi$ are the gravitational accelerations, $\delta v_R$ and $\delta v_\varphi$ are velocity fluctuations, and $\left< \cdot \right>$ denotes an azimuthal average. The vertical integral is calculated by computing $g_R$ and $g_\varphi$ at different values of $z$ \citep[see the Appendix of][for details]{baruteau11}. To compare with equation (\ref{eqabc}), we can use an $\alpha$-parametrisation:
\begin{equation}
\alpha=\left(\frac{d\log \Omega}{d\log R}\right)^{-1}\frac{T_{R\varphi}}{\Sigma \cs^2}.
\end{equation}
We usually find that the disc settles into a state of constant $\alpha$, in which case a single value of $\alpha$ can be assigned to the disc. 

\begin{figure} 
\centering
\resizebox{0.9\hsize}{!}{\includegraphics[]{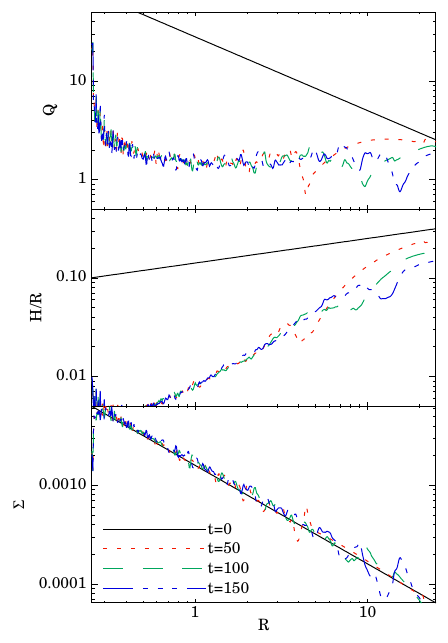}} 
\caption{Radial profile of $Q$ (top panel), $H/R$ (middle panel) and $\Sigma$ (bottom panel) for a resolution $\Delta R/R=0.008$ after 0 (solid curve), 50 (dotted curve), 100 (dashed curve) and 150 (dash-dotted curve) orbits at $R=1$ for $\beta=7.5$.} 
\label{figedge} 
\end{figure} 

\section{Results}
\label{secRes}
At first, the disc is almost axisymmetric with no source of heating. It will therefore start to cool at constant surface density towards $Q=1$. Since $\beta$ is constant, the cooling time scale $t_\mathrm{cool}$ is shortest in the inner parts of the disc, and this is where $Q=1$ is reached first. Therefore, gravito-turbulence starts from the inside and moves outward as more and more of the disc cools towards $Q=1$. At this point, the initial temperature distribution has been washed out completely, and the temperature is now given by the requirement of $Q \approx 1$ with the initial surface density. It is easy to see that this gives $H/R \propto R$ for our initial surface density profile. The number of grid cells per scale height is therefore an increasing function of $R$, which means that the outer parts of the disc are better resolved. The disc then starts to evolve viscously towards a steady-state accreting solution with constant values of  $Q$ and $\alpha$, which has $\Sigma \propto R^{-3/2}$ and constant temperature \citep[see][equation (2.10)]{pringle81}. 

We choose $\beta=7.5$ and evolve the discs for 2000 orbits at $R=1$. According to \cite{rice05} and \cite{gammie01}, these discs should not fragment for $\gamma=5/3$, though previous results may be affected by resolution \citep{meru11}. We increase the resolution to see if convergence can be reached. The resulting surface density distributions are shown in Fig. \ref{figdens}. In our lowest resolution case (left panels), we resolve $H$ outside $R\approx 3$, but we find no evidence of fragmentation (defined such that the fragment is more than 2 orders of magnitude denser than the surroundings). In the final gravito-turbulent state (bottom left panel), we measure a total stress that compares very well with the predicted value of equation (\ref{eqabc}).

When increasing the resolution by a factor of 2 (middle panels of Fig. \ref{figdens}), we do find fragmentation. After 150 orbits at $R=1$, clumps start to form at the outer edge of the turbulent region (top middle panel in Fig. \ref{figdens}). The first surviving clumps form at $R\approx 15$, even though the necessary length scale ($H$) is resolved much further in. When increasing the resolution even further (right panels of Fig. \ref{figdens}), the disc fragments further in ($R\approx 6$). After 300 orbits at $R=1$, several of the fragments have merged or migrated off the computational domain \citep[see][]{baruteau11}, leaving five distinct clumps. 

We have observed fragmentation for values of $\beta$ as high as 15, at a resolution that is 8 times the lowest resolution presented in this Letter. Interestingly, for $\beta=15$, clumps form in the outer disc, migrate inward and leave the computational domain. After all clumps  have left the disc, no further fragmentation is observed. This already suggests that the fragmentation for high $\beta$ is transient and related to the initial state of the disc. 

\begin{figure*} 
\centering
\resizebox{0.9\hsize}{!}{\includegraphics[]{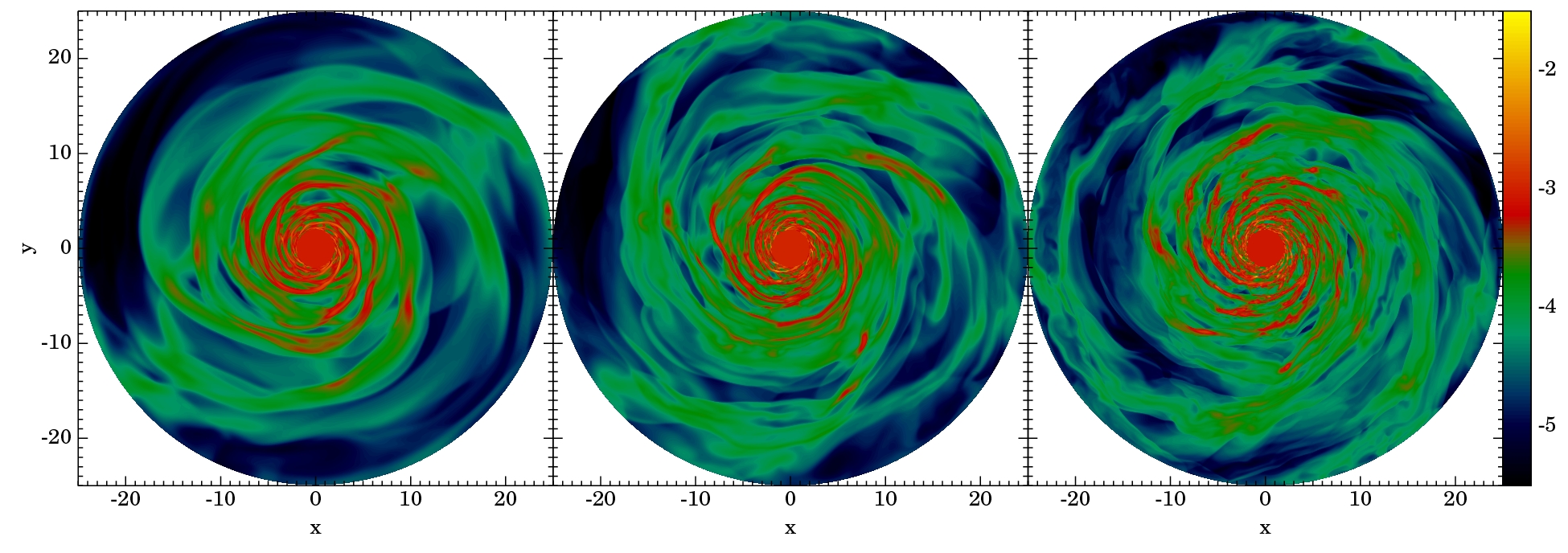}} 
\caption{Surface density after 2000 orbits at $R=1$, where in the first 1000 orbits $\beta=30$ was used, after which $\beta$ was decreased to $7.5$ in the next 500 orbits. From left to right, the resolution is $\Delta R/R=0.008$, $0.004$ and $0.002$.} 
\label{figdens2} 
\end{figure*} 

We can obtain further insight into why these discs fragment by looking at radial profiles of $Q$, $H/R$ and $\Sigma$. Figure \ref{figedge} shows the results for the low resolution simulation with $\beta=7.5$. From the top panel, we see that the disc cools down to a state of constant $Q\approx 1.5$. Note that inside $R=3$, the scale height the disc tries to cool towards can not be resolved at this resolution, so we should only consider $R>3$. As mentioned before, the disc becomes turbulent from the inside out. The outer edge of the turbulent region can be clearly identified as a location where $Q<1$ in the top panel, and as bumps in the temperature and surface density profiles in the bottom two panels. Such a global feature makes the set-up no longer scale-free. A region where $Q<1$ is of course unstable to gravitational instabilities. Moreover, sharp transitions in temperature and surface density are notoriously unstable \citep[e.g.][]{papa85,papa87,lovelace99,linpap11}. It is conceivable that the formation of clumps at the outer edge of the turbulent region, as observed for example in the top middle panel of Fig. \ref{figdens}, is related to this sharp transition between the turbulent and laminar parts of the disc. 

To test to what extent the smooth initial conditions, and the related formation of this global feature, play a part in disc fragmentation, we employ the same strategy as \cite{clarke07}. We start off with a disc that has $\beta=30$, and let gravito-turbulence fully develop over 1000 orbits at $R=1$. For this value of $\beta$, we have not observed fragmentation for any of the resolutions. Over the next 500 orbits, we then linearly decrease $\beta$ to a value of $15, 7.5, 5, 4$ or $2.5$. This way, we avoid any initial transients affecting the possible fragmentation of the disc, and restore the scale-free nature of the set-up. We then evolve the discs with $\beta$ held fixed at the desired values for another 2000 orbits (or until they fragment).

\begin{figure} 
\centering
\resizebox{0.9\hsize}{!}{\includegraphics[]{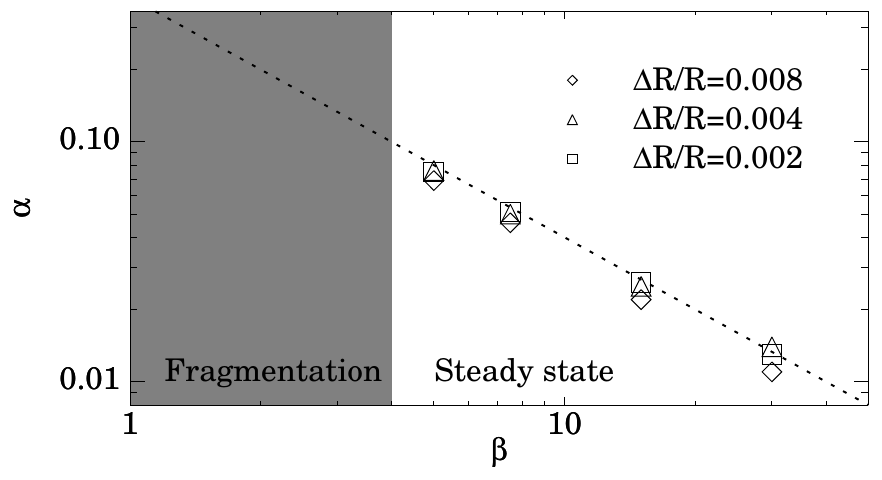}} 
\caption{Measured total shear stress for different final values of $\beta$ at different resolutions. The value of $\alpha$ was calculated by averaging over orbits $1900-2000$ of each run (i.e. when the discs were held at the final constant $\beta$ values), and averaged over the whole disc. The dotted line indicates equation (\ref{eqabc}).} 
\label{figalpha} 
\end{figure}

In Fig. \ref{figdens2}, we show the resulting surface density structures after 2000 orbits for the same three resolutions and the same final $\beta=7.5$ as in Fig. \ref{figdens}. This time, we do not observe fragmentation for any of the resolutions, and the surface densities appear very similar. We also find that the resulting total stresses are equivalent in the three cases (see Fig. \ref{figalpha}), and agree very well with equation (\ref{eqabc}). We therefore conclude that we have reached numerical convergence in this case. A side-effect of letting the disc evolve for 3500 orbits is that mass is redistributed over the computational domain, as the disc is trying to reach a steady state with $\Sigma \propto R^{-3/2}$.

We find that $\beta$ can be reduced to a value of $5$ without seeing fragmentation for all 3 resolutions. From Fig. \ref{figalpha}, we see that in all cases, we find very similar total shear stresses for different resolutions, which in turn agree very well with equation (\ref{eqabc}). For $\beta \lesssim 4$, we find fragmentation for all resolutions considered. The maximum stress the disc can provide to balance the cooling is $\alpha_\mathrm{max} \approx 0.1$. This is in agreement with the results of \cite{clarke07}, who found that the maximum stress increased by approximately a factor of 2 compared to \cite{rice05}, who started from smooth initial conditions. Note that we do not attempt to determine the exact value of $\betac$. Around $\beta=\betac$, other numerical effects may play a role \citep[see][]{clarke07}.

\begin{figure} 
\centering
\resizebox{0.9\hsize}{!}{\includegraphics[]{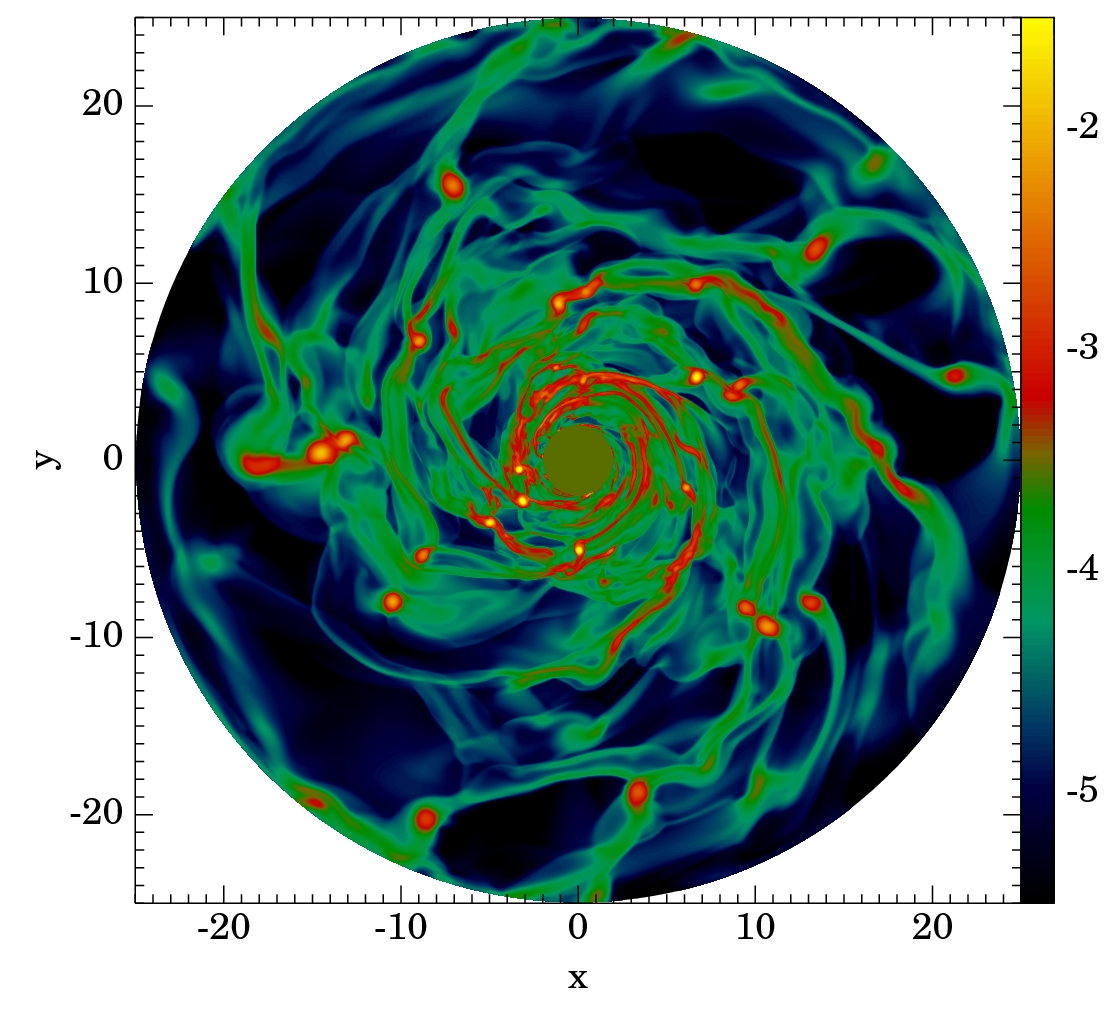}} 
\caption{Surface density after 1550 orbits at $R=1$, of which the first 1000 orbits were run with $\beta=30$, and in the next 500 orbits $\beta$ was decreased to $2.5$. The disc fragmented after 1475 orbits. Resolution is $\Delta R/R=0.004$.} 
\label{figdens3} 
\end{figure} 

Figure \ref{figdens3} shows the fragmented disc with $\beta=2.5$ at $\Delta R/R=0.004$. Note that there are no special locations anymore: the disc fragments everywhere as long as $H$ is resolved, which is basically the criterion of \cite{truelove97}. We also do not find any borderline cases \citep{meru11}, for which the disc fragments at early times but the fragments do not survive. The disc either fragments, after which the formed clumps migrate and merge, or the disc finds itself in a steady gravito-turbulent state. With this setup, a single parameter, $\beta$, determines whether the disc fragments or not.

\section{Discussion and conclusions}
\label{secDisc}
We have studied the numerical convergence of simulations of self-gravitating discs using the grid-based code {\sc fargo}. For smooth initial conditions, we find the same non-convergence as recently reported for SPH simulations \citep{meru11}. 
We have argued that this non-convergence is related to the smooth initial conditions, that lead to an edge in  temperature and surface density at the outer boundary of the gravito-turbulent region. This global feature, which becomes more pronounced at higher resolution, can drive instability at larger values of $\beta$, leading to fragmentation at least for $\beta \leq 15$. It is not clear whether convergence can ever be reached using this set-up. It is beyond the scope of this Letter to characterise these edge effects in detail, because the edge is completely artificial. We comment that for a constant initial surface density, the disc fragments for even higher values of $\beta$ at a given resolution ($\beta=15$ for $\Delta R/R= 0.002$, but note that, since $H/R \propto R^2$ in this case, the number of grid cells per scale height increases rapidly towards the outer disc).
 
It is possible to restore the scale-free nature of the set up by carefully choosing the initial conditions. We have taken the approach of \cite{clarke07}, and first set up a gravito-turbulent disc at a high value of $\beta=30$, for which we do not find fragmentation with the resolutions we consider. We then linearly decrease $\beta$ to the desired value. Using this set up, we find that our results are converged numerically, and the maximum value of $\alpha$ that the disc can sustain, $\alpha_\mathrm{max}\approx 0.1$, is in good agreement with that found in \cite{clarke07}. It is important to note that while \cite{clarke07} interpret their results as being partly dependent on the disc's thermal history, their results may well have been due to the removal of edge effects. We find a critical value of $\beta$, $\betac \approx 4$, in reasonable agreement with \cite{gammie01}. Note also that \cite{gammie01} did not find fragmentation for $\beta=10$ for any of the resolutions he considered, which is not surprising in the light of our findings, since he considered \emph{local} models, from which edges are absent by definition. Note also that at the maximum resolution considered by \cite{gammie01}, $H$ is still only resolved by $\sim$ 10 grid cells. More work is necessary to confirm numerical convergence at even higher resolution, in local as well as global models, where global models should slowly reduce $\beta$ to avoid artefacts from the initial conditions. It is conceivable that at higher resolution, an even slower reduction rate for $\beta$ is necessary to avoid these. Reducing $\beta$ from $30$ to $7.5$ within one orbit at $R=1$ again leads to the formation of an edge and fragmentation similar to the top panels in Fig. \ref{figdens}.

We stress at this point that it may be possible for a disc with $\beta < \betac$ to resist fragmentation (e.g. if other heating mechanisms are present). Likewise, it is interesting to note that if a \emph{physical} edge exists (rather than the artificial numerical one created by smooth initial conditions), it may drive a disc with $\beta > \betac$ towards fragmentation. In this respect, it is not clear what the `correct' initial conditions are, especially since the self-gravitating state occurs shortly after the formation of the disc. Nevertheless, in the idealised case of no edges and a simple cooling law, we have shown that at least spurious fragmentation at high values of $\beta$ can be avoided, and that this result hold for all resolutions considered here.
 
\section*{Acknowledgements}
We thank the referee, Guiseppe Lodato, for an insightful report that helped to improve the clarity of the paper. SJP and CB are supported by STFC and Herchel-Smith postdoctoral fellowships, respectively. FM acknowledges support through grant KL 650/8-2 (within FOR 759) of the German Research Foundation (DFG). Simulations were performed using the Darwin Supercomputer of the University of Cambridge High Performance Computing Service (http://www.hpc.cam.ac.uk), provided by Dell Inc. using Strategic Research Infrastructure Funding from the Higher Education Funding Council for England.

\bibliography{convergence}

\begin{thebibliography}{}

\bibitem[\protect\citeauthoryear{{Balbus} \& {Papaloizou}}{{Balbus} \&
  {Papaloizou}}{1999}]{balbus99}
{Balbus} S.~A.,  {Papaloizou} J.~C.~B.,  1999, \apj, 521, 650

\bibitem[\protect\citeauthoryear{{Baruteau} \& {Masset}}{{Baruteau} \&
  {Masset}}{2008}]{baruteau08}
{Baruteau} C.,  {Masset} F.,  2008, \apj, 678, 483

\bibitem[\protect\citeauthoryear{{Baruteau}, {Meru} \&
  {Paardekooper}}{{Baruteau} et~al.}{2011}]{baruteau11}
{Baruteau} C.,  {Meru} F.,    {Paardekooper} S.-J.,  2011, \mnras, accepted

\bibitem[\protect\citeauthoryear{{Boley}}{{Boley}}{2009}]{boley09}
{Boley} A.~C.,  2009, \apjl, 695, L53

\bibitem[\protect\citeauthoryear{{Boley}, {Hayfield}, {Mayer} \&
  {Durisen}}{{Boley} et~al.}{2010}]{boley10}
{Boley} A.~C.,  {Hayfield} T.,  {Mayer} L.,    {Durisen} R.~H.,  2010, \icarus,
  207, 509

\bibitem[\protect\citeauthoryear{{Boss}}{{Boss}}{1997}]{boss97}
{Boss} A.~P.,  1997, Science, 276, 1836

\bibitem[\protect\citeauthoryear{{Clarke}, {Harper-Clark} \& {Lodato}}{{Clarke}
  et~al.}{2007}]{clarke07}
{Clarke} C.~J.,  {Harper-Clark} E.,    {Lodato} G.,  2007, \mnras, 381, 1543

\bibitem[\protect\citeauthoryear{{Durisen}, {Boss}, {Mayer}, {Nelson}, {Quinn}
  \& {Rice}}{{Durisen} et~al.}{2007}]{durisen07}
{Durisen} R.~H.,  {Boss} A.~P.,  {Mayer} L.,  {Nelson} A.~F.,  {Quinn} T.,
  {Rice} W.~K.~M.,  2007, in {Reipurth} B.,  {Jewitt} D.,   {Keil} K.,  eds,
  Protostars and Planets V {Gravitational Instabilities in Gaseous
  Protoplanetary Disks and Implications for Giant Planet Formation}.
Univ. Arizona press, Tucson, pp 607--622

\bibitem[\protect\citeauthoryear{{Forgan}, {Rice}, {Cossins} \&
  {Lodato}}{{Forgan} et~al.}{2011}]{forgan11}
{Forgan} D.,  {Rice} K.,  {Cossins} P.,    {Lodato} G.,  2011, \mnras, 410, 994

\bibitem[\protect\citeauthoryear{{Gammie}}{{Gammie}}{2001}]{gammie01}
{Gammie} C.~F.,  2001, \apj, 553, 174

\bibitem[\protect\citeauthoryear{{Goldreich} \& {Lynden-Bell}}{{Goldreich} \&
  {Lynden-Bell}}{1965}]{gold65}
{Goldreich} P.,  {Lynden-Bell} D.,  1965, \mnras, 130, 125

\bibitem[\protect\citeauthoryear{{Lin} \& {Papaloizou}}{{Lin} \&
  {Papaloizou}}{2011}]{linpap11}
{Lin} M.-K.,  {Papaloizou} J.,  2011, \mnras, accepted

\bibitem[\protect\citeauthoryear{{Lodato} \& {Clarke}}{{Lodato} \&
  {Clarke}}{2011}]{lodato11}
{Lodato} G.,  {Clarke} C.~J.,  2011, \mnras, in press

\bibitem[\protect\citeauthoryear{{Lodato} \& {Rice}}{{Lodato} \&
  {Rice}}{2004}]{lodato04}
{Lodato} G.,  {Rice} W.~K.~M.,  2004, \mnras, 351, 630

\bibitem[\protect\citeauthoryear{{Lodato} \& {Rice}}{{Lodato} \&
  {Rice}}{2005}]{lodato05}
{Lodato} G.,  {Rice} W.~K.~M.,  2005, \mnras, 358, 1489

\bibitem[\protect\citeauthoryear{{Lovelace}, {Li}, {Colgate} \&
  {Nelson}}{{Lovelace} et~al.}{1999}]{lovelace99}
{Lovelace} R.~V.~E.,  {Li} H.,  {Colgate} S.~A.,    {Nelson} A.~F.,  1999,
  \apj, 513, 805

\bibitem[\protect\citeauthoryear{{Masset}}{{Masset}}{2000a}]{masset00a}
{Masset} F.,  2000a, \aaps, 141, 165

\bibitem[\protect\citeauthoryear{{Masset}}{{Masset}}{2000b}]{masset00b}
{Masset} F.~S.,  2000b, in {G.~Garz{\'o}n, C.~Eiroa, D.~de Winter, \&
  T.~J.~Mahoney} ed., Disks, Planetesimals, and Planets Vol.~219 of ASP Conf.
  Ser., {FARGO: A Fast Eulerian Transport Algorithm for Differentially Rotating
  Disks}.
Astron. Soc. Pac., San Francisco, pp 75--80

\bibitem[\protect\citeauthoryear{{Meru} \& {Bate}}{{Meru} \&
  {Bate}}{2011a}]{meru11}
{Meru} F.,  {Bate} M.~R.,  2011a, \mnras, 411, L1

\bibitem[\protect\citeauthoryear{{Meru} \& {Bate}}{{Meru} \&
  {Bate}}{2011b}]{meru11b}
{Meru} F.,  {Bate} M.~R.,  2011b, \mnras, 410, 559

\bibitem[\protect\citeauthoryear{{Paczynski}}{{Paczynski}}{1978}]{pac78}
{Paczynski} B.,  1978, \actaa, 28, 91

\bibitem[\protect\citeauthoryear{{Papaloizou} \& {Pringle}}{{Papaloizou} \&
  {Pringle}}{1985}]{papa85}
{Papaloizou} J.~C.~B.,  {Pringle} J.~E.,  1985, \mnras, 213, 799

\bibitem[\protect\citeauthoryear{{Papaloizou} \& {Pringle}}{{Papaloizou} \&
  {Pringle}}{1987}]{papa87}
{Papaloizou} J.~C.~B.,  {Pringle} J.~E.,  1987, \mnras, 225, 267

\bibitem[\protect\citeauthoryear{{Pringle}}{{Pringle}}{1981}]{pringle81}
{Pringle} J.~E.,  1981, \araa, 19, 137

\bibitem[\protect\citeauthoryear{{Rafikov}}{{Rafikov}}{2005}]{rafikov05}
{Rafikov} R.~R.,  2005, \apjl, 621, L69

\bibitem[\protect\citeauthoryear{{Rice}, {Lodato} \& {Armitage}}{{Rice}
  et~al.}{2005}]{rice05}
{Rice} W.~K.~M.,  {Lodato} G.,    {Armitage} P.~J.,  2005, \mnras, 364, L56

\bibitem[\protect\citeauthoryear{{Stone} \& {Norman}}{{Stone} \&
  {Norman}}{1992}]{stone92}
{Stone} J.~M.,  {Norman} M.~L.,  1992, \apjs, 80, 753

\bibitem[\protect\citeauthoryear{{Toomre}}{{Toomre}}{1964}]{toomre64}
{Toomre} A.,  1964, \apj, 139, 1217

\bibitem[\protect\citeauthoryear{{Truelove}, {Klein}, {McKee}, {Holliman} II,
  {Howell} \& {Greenough}}{{Truelove} et~al.}{1997}]{truelove97}
{Truelove} J.~K.,  {Klein} R.~I.,  {McKee} C.~F.,  {Holliman} II J.~H.,
  {Howell} L.~H.,    {Greenough} J.~A.,  1997, \apjl, 489, L179

\end{thebibliography}

\label{lastpage}

\end{document}